\documentclass[a4,usenatbib,fleqn]{mn2e}
\usepackage{mathptmx}
\usepackage{amsmath,amssymb,amsbsy}
\usepackage[pdftex]{graphicx}
\usepackage[backref=none]{hyperref}
\usepackage[usenames,dvipsnames]{color}
\usepackage{bm}
\usepackage{txfonts}
\DeclareSymbolFont{cmletters}{OML}{cmm}{m}{it}
\DeclareMathSymbol{v}{\mathalpha}{cmletters}{"76}

\definecolor{MyDarkBlue}{rgb}{0,0.1,0.7}
\hypersetup{pdfborder={0 0 0},colorlinks,breaklinks=true,
  urlcolor={blue},citecolor={MyDarkBlue},linkcolor={magenta}}

\newcommand{\degrees}{^{\circ}}
\newcommand{\msun}{\ensuremath{{\rm M}_{\bigodot}}}

\newcommand{\apj}{ApJ}
\newcommand{\apjl}{ApJ}

\newcommand{\mnras}{MNRAS}
\newcommand{\nat}{Nature}

\newcommand{\aap}{A{\&}A}

\newcommand{\aplett}{Astrophysical Letters}

\newcommand{\cjaa}{Chinese Journal of Astronomy and Astrophysics}
\newcommand{\sovast}{Soviet Astronomy}

\title[On the new braking index of PSR~B0540--69]{On the new braking index of PSR~B0540--69: further support for magnetic field growth of neutron stars following submergence by fallback accretion}

\author[Ek\c{s}i, K.Y.]{K.~Yavuz Ek\c{s}i \\  
 Istanbul Technical University,  Faculty  of Science  and  Letters,  Physics Engineering  Department,
  34469,  Istanbul, Turkey, \href{mailto:eksi@itu.edu.tr}{eksi@itu.edu.tr}
}

\begin{document}
\date{}
\pagerange{\pageref{firstpage}--\pageref{lastpage}} \pubyear{2017}

\maketitle

\begin{abstract}
The magnetic fields of the nascent neutron stars could be submerged in the crust by rapid fallback accretion and could diffuse to the surface later in life. According to this field burial scenario young pulsars may have growing magnetic fields, a process known to result with less-than-three braking indices; larger braking indices implying longer field-growth time-scales. A nascent neutron star with a larger kick velocity would accrete less amount of matter leading to a shallower burial of its field and a more rapid field growth. Such an inverse relation between the field growth time-scale inferred from the braking indices and space velocity of pulsars was claimed in the past as a prediction of the field-burial scenario. With a braking index of $n\sim 2$ and large space velocity PSR~B0540--69 was then an outlier in the claimed relation. The object recently made a transition to a rapid spin-down state accompanied by a low braking index. This new braking index implies a much shorter time-scale for the field growth which is consistent with the high space velocity of the object, in better agreement with the claimed relation. This observation lends support to the field burial scenario and implies that the growth of the magnetic field does not proceed at a constant pace but is slowed or completely halted at times. The slow spin-down state associated with the high braking index before 2011 which lasted for at least about 30 years was then such an episode of slow field growth.
\\
\end{abstract}


\label{firstpage}
\section{Introduction}
\label{sec:intro}

PSR~B0540--69 (hereafter B0540) is a $P = 50~{\rm ms}$ pulsar discovered in X-rays by the Einstein Observatory \citep{sew+84} and then in radio from Parkes \citep{man+93}.  It is located in the Large Magellanic Cloud (LMC) rendering it the first extragalactic pulsar discovered. It is  associated with the
$\sim 1140$-year-old supernova remnant SNR~0540--69.3 \citep{wil+08,bra+14}.
The mass of the progenitor of B0540 is estimated to be in the $20-25\msun$ range \citep{lun+11}. 

B0540 shares similar rotational properties with the Crab pulsar and is sometimes called the `Crab-twin'. It has the third highest  spin-down power, 
$\dot{E}=4{\rm \pi}^2 I \dot{P}/P^3=1.5\times 10^{38}~I_{45}~{\rm erg~s^{-1}}$, among all known rotationally powered pulsars (RPPs); 
here $I_{45}$ is the moment of inertia of the neutron star in units of $10^{45}~{\rm g~cm^2}$. 
B0540 is the most luminous pulsar known in gamma-rays, $L_{\gamma}=7.6 \times
10^{36}~(d/50~{\rm kpc})^2~{\rm erg~s^{-1}}$ \citep{fermi+15} where the distance is scaled to the distance of the LMC known to an accuracy of 2\% \citep{pie+13}. 
Adding the X-ray luminosity \citep{cam+08} the integrated luminosity
of the object is $L_{X+\gamma} \sim 9.7 \times 10^{36}~(d/50~{\rm kpc})^2~{\rm erg~s^{-1}}$ \citep{fermi+15} which is 4 times larger than that of Crab \citep{wil+01}.

Recently, B0540 showed a `large, sudden and persistent' increase of $36\%$ in its spin-down rate $\dot{\nu}$ \citep{mar+15}, accompanied by a dramatic drop in its braking index, $n \equiv \nu \ddot{\nu}/\dot{\nu}^2$, from $n = 2.129\pm 0.012$ \citep{ferd+15} to $n=0.031\pm 0.013$ \citep{mar+16}. This transition has occurred within two weeks in December 2011 after a 30 years of stable slowdown and has persisted from December 2011 through at least June 2016, about $4.5~{\rm years}$. This transition, having a timescale with an upper limit of two weeks, is similar to transitions that occur in other pulsars \citep{lyn+10}, but is not accompanied by any change in the X-ray pulse shape or luminosity.

The observations of \citet{mar+15} and \citet{mar+16} pose two challenges: (i) What causes the change in the spin-down rate and the braking index in such a short time? (ii) what is the cause of such a small measured braking index in the rapid-spin-down state? The explanation for the first question favours changes in the pulsar magnetosphere \citep{li+12a,li+12b}. Several explanations had been given for the origin of the less-than-three braking indices measured from the pulsars some of which could apply for the $n=0.031\pm 0.013$ of B0540 measured in the rapid-spindown state \citep{mar+16}. These include the contribution of propeller torques by supernova debris discs \citep{mic81,men+01,cal+13,ozs+14} or the growth, after initial submergence, of dipole fields by initial rapid accretion of fallback matter soon after the supernova explosion that formed the neutron star \citep{you95,gep+99,esp+11,ho11,vig12,ber+13,ho15,tor+16}. A prediction of the latter view is that the time-scale for the growth of the magnetic field, $\tau_{\mu} \equiv \mu/\dot{\mu}$, is inversely proportional to the space velocity of the pulsar \citep[][hereafter GE13]{gun13} as a neutron star with a large velocity would accrete a less amount of matter---leading to a shallow submergence of its field---from the fallback matter. The time-scale of the growth can be obtained from the measured braking index assuming field growth is the only effect that causes the deviation from the prediction of dipole spin-down with constant magnetic field. 
There are only 5  pulsars with both measured braking indices and space velocities and it is not possible to find a strong relation. 
The work of GE13 yet provided a marginal support \citep{ber+13} for the growth of the submerged fields by claiming such an inverse relation from a very sparse data. The braking index of B0540 employed in GE13 was $n=2.087(7)$ as measured by \citep[][see also \citet{ferd+15}]{gra+11} during the slow-spindown state \citep{mar+15} before the December 2011 transition. This inferred a long time-scale for the field growth although the object is the fastest among all the 5 sources and consequently  B0540 was then an outlier in the claimed relation.  With the recent measurement of the braking index of  B0540 as $n=0.031\pm 0.013$ \citep{mar+16} this object is no longer an outlier but fits in with the claimed relation as the field growth time-scale implied by the new braking index is much shorter. 

The purpose of the present paper is to show that the field growth time-scale inferred from the new braking index measured from B0540 \citep{mar+16} is in better agreement with the relation provided by GE13 as compared to the time-scale inferred from the earlier braking index measurements  obtained in the slow spin-down regime before 2011.
In \autoref{sec:method} the model of magnetic dipole torques in the presence of increasing magnetic moment is introduced. In \autoref{sec:result} the existing data and the result from the model are presented. 
In \autoref{sec:discuss} the implications of these results are discussed.

\section{Method}
\label{sec:method}

\begin{table*}
\begin{minipage}{160mm}
\caption{Input parameters of the pulsars with accurately measured braking indices and space velocities.
	\label{tbl:data}}
\centering	
\begin{tabular}{lccrlc}
\hline
Pulsar            & $\nu$     & $\tau_{c}$      & $n$ \phantom{xxxxx}  &  $v$ \phantom{xxx}  &  $\tau_{\mu}$    \\
                  & (Hz)      & (kyr)           &                      & (km s$^{-1}$)       &   (kyr)              \\
\hline 
B0531$+$21(Crab) & $29.947$    & $1.257$    & $2.342(1)$ [1]   & $140 \pm 8$ [2]  & $6.59\pm 1.06$     \\
B0833--45(Vela)  & $11.2(5)$   & $11.30$    & $1.7(2)$ [3]        & $61 \pm 2$  [4]  & $27.3\pm 7.63$       \\
J1833--1034      & $16.159$    & $4.854$    & $1.8569(6)$ [5]     & $125 \pm 30$ [4] & $15.46\pm 1.53$   \\
J0537--6910      & $62.018$    & $4.93$     & $-1.2$ [6]        & $634 \pm 50$ [4]   & $4.60 \pm 0.45$   \\
\hline
B0540--69 (pre-tr)  & $19.80$    & $1.67$    & $2.129\pm 0.012$ [7]    & $1300\pm 612$ [4]   & $6.84\pm 0.94$     \\
B0540--69 (post-tr) & $19.70$    & $1.23$    & $0.031\pm 0.013$ [8]    & $ 1300\pm 612$ [4]   & $1.59 \pm 0.07$     \\
\hline
\end{tabular}
\medskip  \\
Numbers in parenthesis are the last digit errors. The calculation of $\tau_{\mu}$ is explained in the text.
[1] \citet{lyn+15}; note that in GE13 we have employed $n=2.51(1)$ given by \citet{lyn+93}.
[2] \citet{ng06}, see \citet{kap+08} cited in \citet{han+15} for a velocity of 120~km~s$^{-1}$ for Crab pulsar,
[3] \citet{esp+17}; note that in GE13 we have employed $n=1.4$ given by \citet{lyn+96}; note also that \citet{akb+16} gives $ n = 2.81 \pm 0.12$ for this object,
[4] \citet{ng07}, 
[5] \citet{roy+12}, 
[6] Antonopoulou et al.\ in preparation as cited in  \citet{esp+17}. Although an error is not given we assumed $\Delta n = 0.3$ for this object. Note that in GE13 we have employed $n=-1.5$ as given by \citet{mid+06}.
[7] \citet{ferd+15},
[10] \citet{zha00},
[8] \citet{mar+16},
[12] \citet{hal+09}.

\end{minipage}
\end{table*}

Fallback accretion  after a supernova explosion onto newborn neutron stars is expected on very general grounds \citep{col71,zel72,che89}.
The submergence of the magnetic field of a nascent neutron star by fallback accretion is an idea dating back to 90's \citep{mus95,you95,mus96,gep+99}. 
The interest to the model is recently rekindled \citep{ho11,vig12,ber+13,tor+16,rog16} by the identification of central compact objects \citep[CCOs: see][for a review]{del08} 
which appear to have small \textit{dipole} magnetic fields \citep[e.g.][]{got+13}, but likely to have substantial subsurface magnetic fields as implied by the anisotropic temperature distribution on their surface \citep{sha12,bog14}. Another motivation for the model is the very small braking index of PSR J1734--3333 which can be explained by the growth of the submerged field \citep{esp+11,ho15}. 

The Ohmic time-scale  for the diffusion of the magnetic field back to the surface is $\tau_{\rm Ohm}=4{\rm  \pi} \sigma (\Delta R)^2/c^2$ where $\Delta R$ is the depth at which the magnetic field is buried and $\sigma$ is the conductivity of the crustal matter. The conductivity depends strongly on the depth setting the regime for the conductivity i.e.\ whether electron-electron scattering, electron-phonon scattering or impurity scattering dominates. As a result the diffusion time-scale has a strong dependence on the depth. The amount of mass that has accreted determines the depth of the burial \citep{gep+99,ho11,tor+16}.
The mass of crustal matter, $\Delta M$, is related to the depth as $\Delta M \propto (\Delta R)^{1/4}$ (see GE13 based on \citet{lor+93} and \citet{urp79}). 
Neutron stars usually have large space velocities exceeding that of their progenitors, possibly due to asymmetry in the supernova explosion.
Assuming Bondi-Hoyle accretion occurs onto the star from the fallback matter one would expect the accretion rate to be inversely proportional to the cube of the velocity of the neutron star with respect to the fallback matter, $\dot{M} \propto v^{-3}$. A detailed analysis of the geometry of accretion is given in \citet{zha+07} and the implications of the presence of turbulence in the medium is studied by \citep{kru+06}. Detailed simulations are performed by \citet{tor+12}. GE13,  assuming Eddington limited accretion and electron-phonon scattering to dominate the electron conduction found  $\tau_{\rm Ohm} \propto v^{-1}$. It is, however, likely that the energy released by accretion will be emitted by neutrinos so that Eddington limit is not relevant \citep[e.g.][]{hou91}.  Given that there are many uncertain elements in the physics of the accretion and the conductivity, and the impossibility of constraining detailed models with the existing small number of data points, we do not attempt to calculate a detailed relation for $\tau_{\rm Ohm}(v)$, but only argue that $\tau_{\rm Ohm}$ should drop with the space velocity, $v$; higher the space velocity, lower is the accreted mass, shallower is the burial and lower is the conductivity.

We assume that RPPs spin-down under the magnetic dipole torque. This torque has a spin-down component as well as a component leading to the alignment of the magnetic and rotating axis \citep{mic70,dav70}. 
The pulsar is expected have a corotating plasma formed by charged particles ripped off by the strong electric fields from the surface of the neutron star \citep{gol69}. In the presence of a corotating plasma the spin-down  torque is 
\begin{equation}
I \dot{ \Omega} = -\frac{ \mu^2 \Omega^3}{c^3} (1 + \sin^2 \alpha ) 
\label{mdr1} 
\end{equation}
\citep{spi06} and the alignment torque is
\begin{equation}
I \Omega \dot{\alpha} = -\frac{ \mu^2 \Omega^3}{c^3} \sin \alpha \cos \alpha 
\label{mdr2}
\end{equation}
\citep{phi+14}. These are obtained by fitting the results of detailed numerical simulations. 
Here $\Omega=2\pi/P$, $\mu$ and $I$ are the angular velocity, the magnetic moment and moment of inertia of the star, respectively, and $\alpha$ is the inclination angle between rotation and magnetic axis; $c$ is the speed of light. Note that the alignment component of the magnetic dipole torque was not employed in GE13 as is customary in most work on pulsars. The torque is a vector quantity and the alignment torque is an intrinsic component of it \citep{phi+14}. The recently measured greater-than-three braking index of PSR J1640--4631 \citep{arc+16} can be explained by the existence of this torque \citep{arz+15,eks+16} or by the decay of the magnetic field \citep{ho11,ho15}.

We assume that the magnetic dipole moment $\mu$ can change in time; in the case of young pulsars it may increase as a result of diffusion to the surface after being submerged in the crust. In this case the braking index is obtained as
\begin{equation}
n = 3 + 2 u^2 - 4 \frac{\tau_{\rm c}}{\tau_{\mu}}
\label{brake}
\end{equation}
where $\tau_{\rm c} \equiv P/2\dot{P}$ is the characteristic age, $\tau_{\mu} \equiv \mu/\dot{\mu}$ is the field growth time-scale and
\begin{equation}
u = \frac{\sin \alpha \cos \alpha}{1 + \sin^2 \alpha}.
\end{equation}

\begin{figure*}
\centering
\includegraphics[width=0.65\textwidth]{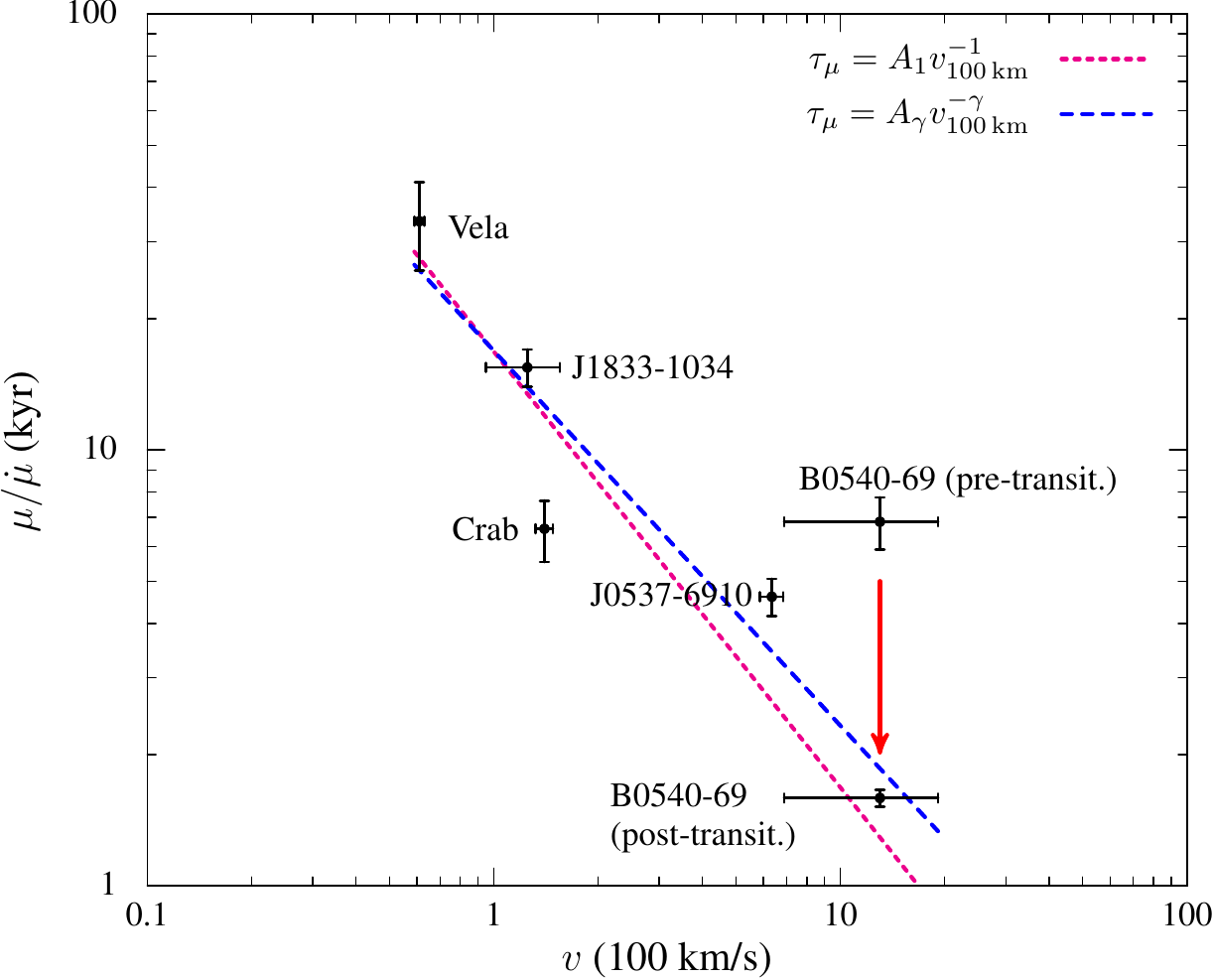}
\caption{The relation between the measured space velocities and
characteristic magnetic field growth time-scale $\tau_{\mu}$ inferred from
measured braking indices via \autoref{taum}. The arrow shows the transition of B0540 from the previously inferred value to the value inferred in this work relying on the measurement of \citep{mar+16} in the rapid-spin-down episode. The long dashed line (coloured blue in the electronic version) is the best fit obtained by \autoref{tau_mu} with both $A_{\gamma}$ and $\gamma$ free; the short dashed line (magenta in the electronic version) is the best fit obtained while $\gamma$ is fixed ($\gamma=1$) as implied by the theory (GE13).
The new position of the object in the diagram lends support for the claimed correlation between the field growth time-scale and the space velocity of the pulsar.}
\label{fig:tauv}
\end{figure*}
If the braking index and the inclination angle are measured for a pulsar one can infer from \autoref{brake} the time-scale for the growth of the magnetic field as
\begin{equation}
\tau_{\mu} = \frac{4\tau_{\rm c}}{3 + 2 u^2 - n}.
\label{taum}
\end{equation}
Here the term $2u^2$ ranges between $0$ (attained at angles $0\degrees$ and $90\degrees$) and $0.25$ attained at $35.25\degrees$. In case $n$ is greater than $3 + 2 u^2$ one would obtain a negative $\tau_{\mu}$ which would imply a decaying magnetic field.

We have compiled the data for the measured braking indices and velocities determined from the proper motion in \autoref{tbl:data}. We calculated $\tau_{\mu}$ from \autoref{taum}.
There are two different uncertainties on the value of $\tau_{\mu}$: due to the measurement error in the braking index and due to the uncertain inclination angle determining $2u^2$. We took into account both uncertainties. Given that inclination angle $\alpha$ of some of the objects are not measured, and that measured ones have very different values, the range of $\tau_{\mu}$ for a pulsar is calculated as follows: The minimum (maximum) value of $\tau_{\mu}$ for a pulsar is obtained by choosing $\alpha=35.25\degrees$ ($\alpha=0\degrees$) corresponding to $2u^2= 0.25$ ($2u^2=0$) and the lower (upper) limit of the measured braking index.

\section{Results}
\label{sec:result}

We plot the data given in \autoref{tbl:data} in \autoref{fig:tauv}. We see that although the number of data is very small, an inverse relation between the velocity and the field-growth time-scale can be judged lending a marginal support to the field burial scenario.
We see that the drop in the braking index of B0540 results with a much shorter time-scale for the field growth as would be expected for such a high speed object. 
The Spearman coefficient for the correlation in the data set before the transition of this object is $\rho = -0.7$
while it is $\rho=-1$ after the transition.
By normal standards, the association between the two variables would be considered statistically significant in both cases; $\tau_{\mu}$ obtained with the braking index measured after transition to the rapid-spindown episode, however, is significantly in better agreement with the expected trend as can be seen from \autoref{fig:tauv}.

We have fitted the data (after the transition to low-braking index state) with
\begin{equation}
\tau_{\mu} = A_{\gamma} \left( \frac{v}{100~{\rm km~s^{-1}}} \right)^{-\gamma}
\label{tau_mu}
\end{equation}
and found $A_\gamma   = 16.88$~kyr and $\gamma = 0.86$. 
We have also fixed $\gamma = 1$ as implied by the theoretical calculations of GE13 and fitted the data with $\tau_{\mu} = A_1 (v/100~{\rm km~s^{-1}})^{-1}$ to find
$A_1 = 16.8 \pm 1.0 $~kyr.

\section{Discussion}
\label{sec:discuss}

An evidence is provided for the existence of a relation between  the space velocity of pulsars and their magnetic field growth time-scales as inferred from the measured braking indices. 
Such a relation would be expected, as first suggested by GE13, within the field burial scenario \citep{gep+99}. 
If the determined relation is not by chance (this is still possible given the number of data points) and in lack of any other motivation for why the braking index of a pulsar would be related to the kick velocity it acquires in the supernova explosion, the result supports the field burial scenario from which such a relation is predicted (GE13). 
Many factors such as the initial conditions of the fallback accretion (i.e.\ $\rho_0 t_0^3$ in GE13) and uncertainties in the measurements of the velocities and the braking indices could affect the results. 

The change in the braking index of B0540 \citep{mar+16}, according to the arguments in this work, implies that the object's magnetic field has entered into an episode of enhanced growth as is the case with other pulsars employed in this work. This transition occurred  after remaining at the slow growth state for at least about 30 years since the object was discovered \citep{sew+84}. This implies that the growth of the magnetic field of young pulsars does not proceed at a constant pace but might be slowed down at times. The high braking index state of B0540 before December 2011 was then such an episode of halted field growth. The transition to the rapid growth state may indicate the breaking of the crust \citep{gou15} yet we note that no accompanying glitch with the transition is reported \citep{mar+16}.

Of the five objects with both measured breaking index and space velocity four of them were in the rapid growth state obeying the relation suggested by GE13 before 2011. That all five objects now appear to be in the rapid growth state implies that the slow field growth states occupy a shorter time in the lifetime of an individual young pulsar. 

It would be very interesting to see whether the objects with measured braking indices but without a measurement of space velocity, namely PSRs B1509--58, J1846--0258, J1119--6127 and J1734--3333 and J1640--4631 obey the claimed relation.

It is tempting to think within the field burial scenario that CCOs, which have two orders of magnitude smaller dipole magnetic fields compared to RPPs, have their fields buried so deeply that their fields could not have the time to diffuse back to the surface. Yet the measured space velocities of CCOs \citep{bec+12,hal15} do not indicate that they have small space velocities as a group. According to the velocity versus field growth time scale relation proposed in this work we can not thus expect them to have especially deep buried fields that grow at a very long time scale. Instead, the measured high velocities imply that they likely have  shallow  buried magnetic fields growing rapidly. The small dipole fields of CCOs do not lead to a sufficiently high spin-down rate that could allow for the measurement of their braking index. It is, thus, not possible to infer the rate of growth of the magnetic fields of CCOs within the framework of the present study. The lack of pulsars, in the observed sample, that seem to have evolved from CCO's \citep{bog+14,luo+15} suggest  that the magnetic fields of CCOs grow rapidly. The rapid growth of the fields of CCOs may address the lack of descendants in the vicinity of CCOs in the $P-\dot{P}$ diagram. Recent studies of the evolution of multipolar magnetic fields \citep{igo+16} suggest that for weak field strengths, such as in the case of CCOs, the higher multipoles re-emerge in a shorter time-scale than the dipole field. This could explain the strong surface fields evidenced by high pulsed fraction \citep{sha12}.

\section*{Acknowledgements}
I thank Erbil G{\"u}gercino{\u g}lu, Sinem {\c S}a{\c s}maz Mu{\c s} and Wynn Ho for reading the earlier version of the manuscript and for useful suggestions. I thank the anonymous referee for useful comments.

\footnotesize{

}

\label{lastpage}

\end{document}